\documentclass[pdflatex,sn-mathphys-num]{sn-jnl}

\usepackage{tabularx}
\usepackage{graphicx}%
\usepackage{multirow}%
\usepackage{amsmath,amssymb,amsfonts}%
\usepackage{amsthm}%
\usepackage[title]{appendix}%
\usepackage{xcolor}%
\usepackage{textcomp}%
\usepackage{manyfoot}%
\usepackage{booktabs}%
\usepackage{algorithm}%
\usepackage{algorithmicx}%
\usepackage{algpseudocode}%
\usepackage{listings}%

\theoremstyle{thmstyleone}%
\theoremstyle{thmstyletwo}%

\theoremstyle{thmstylethree}%

\usepackage{xcolor}

\begin{document}

\title[Persistent Structural Inequality of Online Interactions Across Platforms]{Persistent Structural Inequality of Online Interactions Across Platforms}


\author*[1]{\fnm{Giulio} \sur{Pecile}}\email{giulio.pecile@uniroma1.it}

\author[2]{\fnm{Edoardo} \sur{Di Martino}}\email{edoardo.dimartino@uniroma1.it}

\author[3]{\fnm{Edoardo} \sur{Loru}}\email{edoardo.loru@uniroma1.it}

\author[2]{\fnm{Simon} \sur{Zollo}}\email{simon.zollo@uniroma1.it}

\author[4]{\fnm{Niccolò}
\sur{Di Marco}\email{niccolo.dimarco@unitus.it}}

\author[1]{\fnm{Matteo} \sur{Cinelli}}\email{matteo.cinelli@uniroma1.it}

\affil*[1]{\orgdiv{Department of Computer Science}, \orgname{Sapienza University of Rome}, \orgaddress{\street{Viale Regina Elena 295}, \city{Rome}, \postcode{00161}, \country{Italy}}}

\affil[2]{\orgdiv{Department of Social Sciences and Economics}, \orgname{Sapienza University of Rome}, \orgaddress{\street{P.le Aldo Moro 5}, \city{Rome}, \postcode{00185}, \country{Italy}}}

\affil[3]{\orgdiv{Department of Computer, Control and Management Engineering}, \orgname{Sapienza University of Rome}, \orgaddress{\street{Via Ariosto 25}, \city{Rome}, \postcode{00185}, \country{Italy}}}

\affil[4]{\orgdiv{Department of Legal, Social, and Educational Sciences}, \orgname{Tuscia University} \orgaddress{\street{Via S.M. in Gradi n.4}, \city{Viterbo}, \postcode{01100}, \country{Italy}}}


\abstract{User interactions on social media platforms are unevenly distributed: a small subset of users consistently captures most of the activity, while the majority remains marginal. Although this pattern is well known and often described by power-law distributions, its consistency across time, platforms, and interaction types has not been systematically assessed.
In this study, we analyze user–post bipartite networks from multiple social media platforms. We consider both active contributions (posts) and passive engagement (likes and comments), and quantify distributional properties and inequality using a KL-divergence–based model comparison, an inverse coefficient of variation, and a log-transformed Gini index.
Our results show that interaction inequality remains stable over time within each platform. 
This holds across systems with different sizes, topical focuses, and governance models. These findings indicate that inequality in online engagement is not incidental but reflects structural constraints that shape how visibility and participation are distributed in digital environments.}

\keywords{social media, interaction patterns, interaction distribution, online behavior, interaction inequality, attention dynamics}



\maketitle

\section{Introduction}

The growth of social media has signified that the online domain has become an ever more dominant setting for public debate \cite{DiMarco2026patterns}. This debate usually takes the form of an initial user publishing a post, and others replying or reacting to the message. Because of the online nature of these dynamics, users have access to much larger audiences, which, in turn, can interact with a wider range of posts and opinions. While this has increased the ability to participate in public debates, it has also fostered many unintended phenomena, including the spread of disinformation~\cite{diaz2023disinformation, Budak2024_misinfo} and toxic debate~\cite{avalle2024persistent, tahmasbi2021go, lorutox, falkenberg24patterns}, and the emergence of polarization~\cite{bail2024exposure, kubin2021role, pecile2025, falkenberg23affective, Loru2025agenda}. It is thus important to understand how responsibility for these phenomena is spread across its users. In the case of disinformation, for example, findings suggest that an extremely small group is responsible for very high exposure of misinformation~\cite{disinfo12, Budak2024_misinfo}, particularly under information void regimes ~\cite{scalco2026detect}. 
In particular, a well-documented feature of digital platforms is the unequal distribution of attention: a small share of users consistently attracts most interactions, while the majority remains marginal~\cite{adamic2000power,muchnik2013origins,asur2011trends,ross2015understanding,mathews2017nature}. This pattern recurs across platforms with different user bases, content types, and algorithmic infrastructures, suggesting the presence of structural regularities \cite{di2024patterns,avalle2024persistent}. Power-law and Zipf-like models are often used to describe these dynamics \cite{robins2009closure,ahn2007analysis}, yet it remains unclear whether the observed inequality arises from platform-specific factors \cite{agichtein2008finding,zafarani2014social} or from more general constraints on how attention is distributed online \cite{cinelli2020selective,sangiorgio2025evaluating, DiMarco2024_volatility, zollo2026examining}. This uncertainty challenges the idea that design-level interventions alone can reshape interaction patterns: if inequality consistently re-emerges under different configurations, it may reflect deeper constraints inherent to digital systems. Some work has examined inequality in user activity on platforms such as Reddit~\cite{machado2025super, Panek2018, singer14} and X~\cite{ORELLANARODRIGUEZ201874, zhu2016attention, Bagdouri_2021, Rodriguez17}, but a systematic account of temporal and cross-platform regularities is still missing. Most studies focus on single platforms or isolated events, making it difficult to systematically assess whether observed inequalities reflect transient conditions or more persistent constraints. 

In this work, we examine whether interaction inequality follows consistent trajectories over time and across platforms. In particular, we test whether the emergence of dominant contributors, users that attract disproportionate attention, is a transient artifact or a persistent structural feature. We adopt a network-based approach, modeling user–post interactions and tracking their evolution across time windows. The analysis covers both mainstream platforms and alternative ecosystems, which differ widely in terms of governance, user demographics, and content policies. By examining the concentration and scaling properties of attention across these contexts, we assess not only the internal dynamics of each platform but also whether structural regularities persist across diverse socio-technical environments.

Our approach moves beyond a platform-specific focus to test whether interaction inequality reflects broader constraints linked to limited visibility, attention bottlenecks, and platform-mediated exposure. This perspective presents online engagement as a measurable pattern shaped by the architecture of participation, rather than an outcome of content or ideology. To this end, we adopt a set of complementary indices to capture functional form, tail dominance, and concentration.

\section{Materials and methods}
\subsection{Data}
We analyze a broad dataset covering multiple social media platforms that differ in user experience, governance, and user base composition. While some platforms do allow users to perform other actions besides the ones in our analysis, our work is limited by the ones that were available to us. We also performed a selection of the considered type of interactions to have a coherent analysis across platforms. Among the mainstream platforms, we include X, where posts receive likes and replies; and YouTube, where videos receive likes and comments. We also include Gab, where posts receive favourites and replies. Notably, this platform stands in stark contrast to the others as an alternative network with significantly looser moderation policies. As some platforms also vary in levels of homogeneity of type of discourse, when possible, we also used multiple datasets from the same platforms, representing both casual conversations and politically relevant discussions. Below, we outline the data collection procedure for each.

For what concerns mainstream platforms, we used posts on X that discussed the 26th Conference of the Parties, referred to as COP26 in this work~\cite{falkenberg2022growing}, and the Russo-Ukranian conflict \cite{Chen_Ferrara_2023}, created in 2021 and between 2022 and 2023 respectively. The COP26 dataset contains 11.7 million posts and the Ukraine dataset consists of 103.9 million posts. We also include two X datasets containing casual conversations: ``X GoT'', with 218 thousand posts discussing the TV series \textit{Game of Thrones}, and ``X NASA'', with 151 thousand posts discussing the space agency.
YouTube data were collected using the platform's public API and consist of two datasets covering current events and political discussions, and a third one discussing football. Data from Gab~\cite{dimartino2025} were collected in two stages. Between June 1 and October 23, 2020, posts were retrieved through the general stream. After the API was deprecated, collection continued from the timelines of 930,000 previously identified users. Applying filters for the keywords ``Trump'' and ``Biden'' produced a subset of 468,000 posts covering the period from June to December 2020.

\begin{figure}[!ht]
  \centering
  \includegraphics{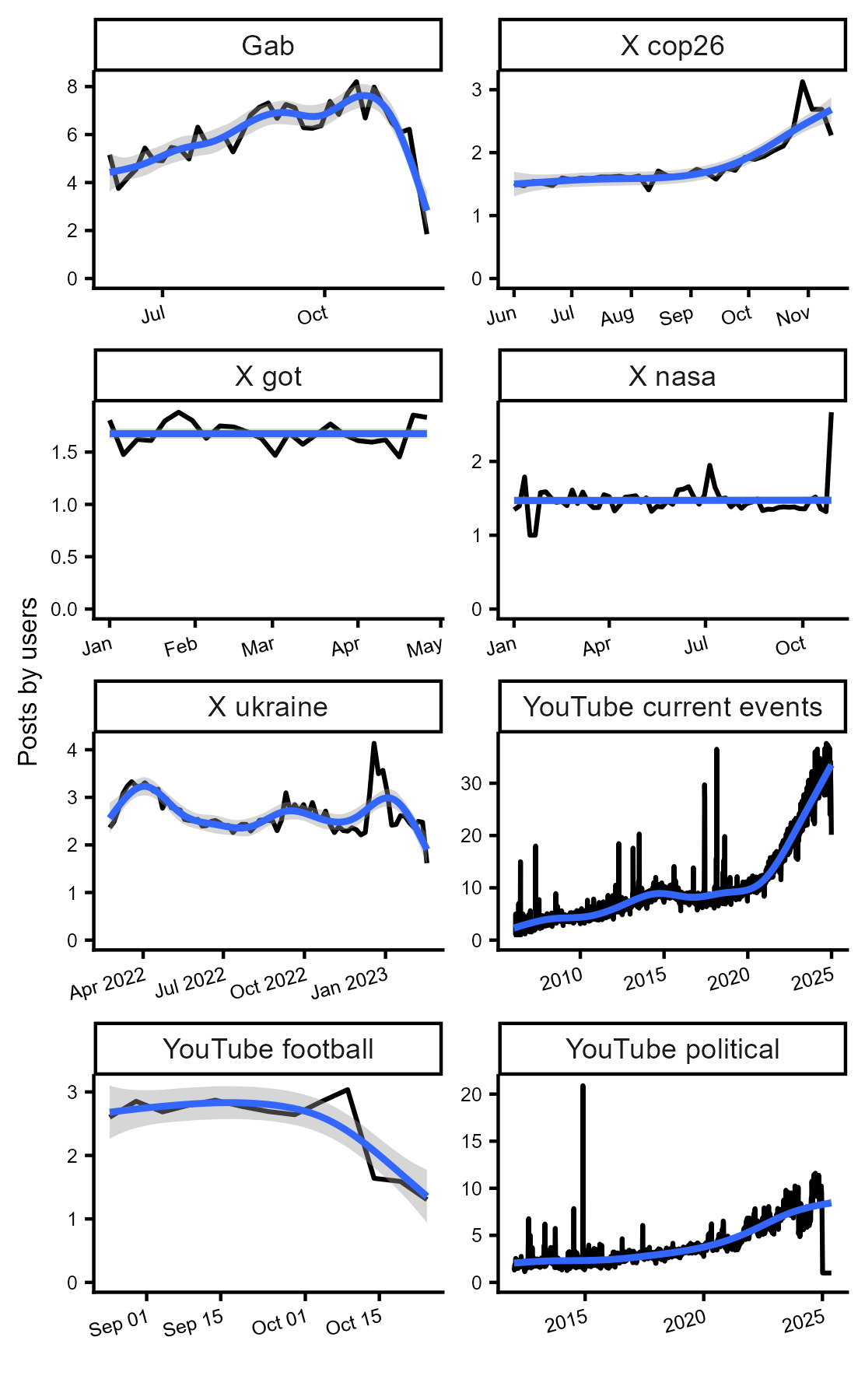}
  \caption{\textbf{Average number of posts created by users.} Almost all platforms show a very limited variation in the average number of posts created by users in the period of analysis. ``YouTube current events'' and ``YouTube political'' are the exceptions, displaying a significant growth over the years.}
  \label{fig:new_users}
\end{figure}

To ensure consistency across platforms, in this work we adopt the terms ``posts'', ``likes'', and ``comments'' for all forms of content and interaction, regardless of platform-specific terminology. For each post, we retain the author, timestamp, and aggregated engagement metrics. A full summary of the platforms, timespans, and original reactions is shown in Table~\ref{tab:datasets}. The datasets vary significantly in size, topic, and temporal coverage, ranging from hundreds of thousands to over 100 million posts, and spanning different time periods. To enable comparative analysis, we discretize time into non-overlapping windows, adjusting the granularity to match each dataset’s volume and timespan. Figure~\ref{fig:new_users} shows the average number of posts created by users during each time window. We observe that in all platforms this value is remarkably stable: the only platform that exhibits a considerable change in this value is YouTube. However, this platform has been in existence for much longer, becoming more mainstream through time.  

\begin{table}[!ht]
\centering
\caption{Data breakdown for all social media platforms considered. Note how for YouTube, ``users'' refers to channels.}
\label{tab:datasets}
\begin{tabularx}{\linewidth}{lllrrX}
\toprule
Platform & From & To & Posts & Users & Reactions \\
\midrule
Gab & 06/2020 & 11/2020 & 468k & 14k & replies, reblogs, favourites \\
X (GOT) & 01/2019 & 04/2019 & 434k & 218k & likes, retweets, replies \\
X (NASA) & 01/2019 & 10/2019 & 151k & 77k & likes, retweets, replies \\
X (COP26) & 06/2021 & 11/2021 & 10M & 2M & likes, retweets, replies \\
X (Ukraine) & 02/2022 & 02/2023 & 87M & 7M & likes, retweets, replies \\
YouTube (political) & 01/2012 & 05/2025 & 278k & 319 & likes, comments \\
YouTube (football) & 08/2022 & 10/2022 & 16k & 1673 & comments \\
YouTube (current events) & 01/2006 & 12/2024 & 12M & 2932 & likes, comments \\
\bottomrule
\end{tabularx}
\end{table}

\subsection{Measures}\label{sec:measures}
In this work we use multiple measures and indices to characterize the levels of concentration in users' interactions. Before computing any measure we divide each dataset in discrete, separate time windows, each of length 5 days. Then, for each user, we compute a summary statistic by aggregating post-level values. Specifically, for each window, we take the median of a chosen post-level quantity (e.g., number of comments or likes), and round it up when the user has an even number of posts. Namely, if $v_{i}$ denotes the post-level quantity of post $i$ in the window, we compute:

\begin{equation*}
m = \lceil \text{median}(v_{1}, v_{2}, \dots) \rceil
\end{equation*}

\begin{figure}[!ht]
  \centering
  \includegraphics[width = \linewidth]{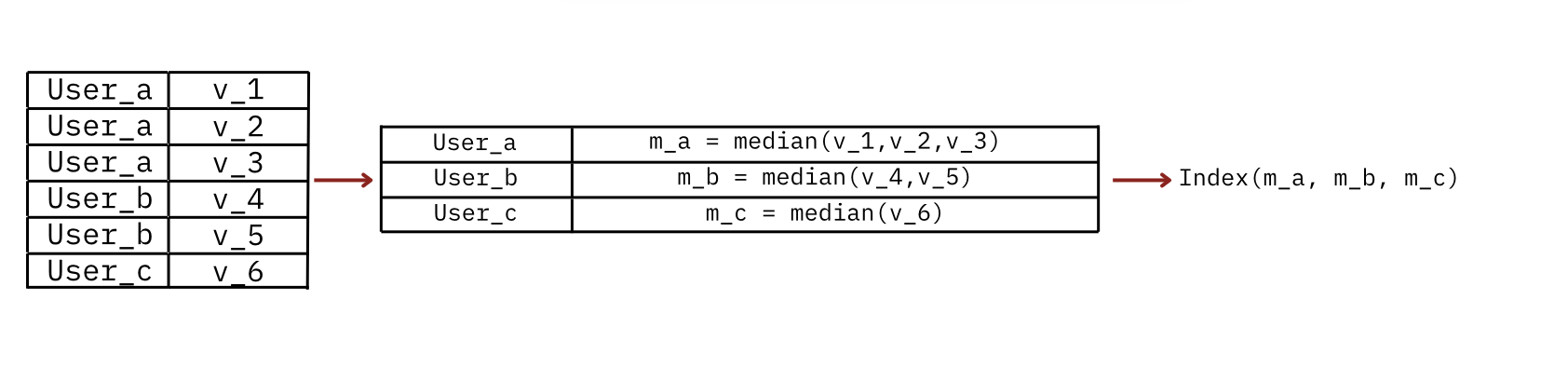}
  \caption{\textbf{Post-level statistics are grouped by users and summarized using the median.}}
  \label{fig:index_process}
\end{figure}

This choice allows us to analyse structural concentration with measures that are robust to outliers. In fact, users who consistently have very high levels of interactions are not mixed with others.
The full procedure is illustrated in Figure~\ref{fig:index_process}.

The three measures capture complementary dimensions of the distribution, global shape (KL ratio), tail dominance (inverse coefficient of variation), and overall concentration (log-Gini). Since we consider the trends of these measures across time, we perform the Mann-Kendall test, which assesses monotonic trends in the time series. We provide the $\tau$ statistic and a p-value in tables presented in the Results. To aid readability, we present the measures in the following paragraphs. 

\subsubsection*{KL ratio} 
The first characteristic we investigate is the shape of the distributions. To this end, we adapt a method introduced by Clauset et al.~\cite{clauset.2009}. For each time window, we fit both an exponential and a power-law distribution to the empirical data. We then compute the Kullback--Leibler divergence between the empirical distribution $P$ and each fitted model, denoted as $Q_{\mathrm{exp}}$ and $Q_{\mathrm{pl}}$, respectively.

We define the divergence ratio as
\begin{equation}\label{eq:ratio}
    R_{KL} = \frac{D_{KL}(P \,\|\, Q_{\mathrm{pl}})}{D_{KL}(P \,\|\, Q_{\mathrm{exp}})}.
\end{equation}

Values of $R_{KL} < 1$ indicate that the empirical distribution is better approximated by a power-law model, whereas $R_{KL} > 1$ suggests a closer fit to an exponential distribution. This comparison is particularly informative, as the two models exhibit markedly different tail behaviors.

\subsubsection*{Inverse Coefficient of Variation}
Next, we further characterize the role of the tail using the inverse coefficient of variation (ICV), defined as the ratio between the mean and the standard deviation:
\begin{equation}\label{eq:ICV}
    \text{ICV} = \frac \mu \sigma.
\end{equation}

The interpretation of ICV is straightforward: higher ICV values indicate that the distribution is more homogeneous, with observations tightly clustered around the mean. Conversely, lower ICV values signal greater dispersion and heterogeneity, often reflecting the presence of heavy tails or extreme values.
For an exponential distributions this ratio is equal to $1$. In contrast, heavy-tailed distributions such as power-laws typically exhibit $\sigma \gg \mu$, leading to ICV $\ll 1$. Therefore, smaller values of ICV indicate a stronger dominance of the tail and a higher level of dispersion.

\subsubsection*{Gini Index}
Once the general shape of the distribution and the role of its tail have been characterized, we move to quantifying the level of inequality using the Gini index. 

The Gini index is a widely used measure of inequality that quantifies how unevenly a quantity is distributed across a population. It ranges from 0 to 1, where 0 indicates perfect equality (all observations have the same value) and 1 corresponds to maximal inequality (all the mass is concentrated in a single observation). In our context, it captures how unevenly activity (e.g., posts, reactions) is distributed across users.
Mathematically, given a set of non-negative activity values $x_1, x_2, \dots, x_n$ with mean $\mu$, the Gini index is defined as
\begin{equation*}\label{eq:loggini}
    G = \frac{1}{2 n^2 \mu} \sum_{i=1}^{n} \sum_{j=1}^{n} |x_i - x_j|.
\end{equation*}

However, in the presence of extremely heavy-tailed distributions, this metric tends to saturate near 1, limiting its ability to resolve differences. To address this limitation, we compute the Gini index on a log-transformed variable, $Y = \log (X +1)$. This transformation compresses the scale of large values while preserving the ordinal structure of the data, allowing for a more informative comparison across highly skewed distributions. We observe that distributions that have an index equal to 0 or 1 in the original space are also mapped to observations with indices equal to 0 and 1.
Importantly, observing high inequality even in log-transformed space indicates that concentration reflects a structurally imbalanced distribution and is not just driven by extreme outliers. 

\section{Results and Discussion}
\subsection{Shape of distribution}
\begin{figure}[!ht]
  \centering
  \includegraphics[width = \linewidth]{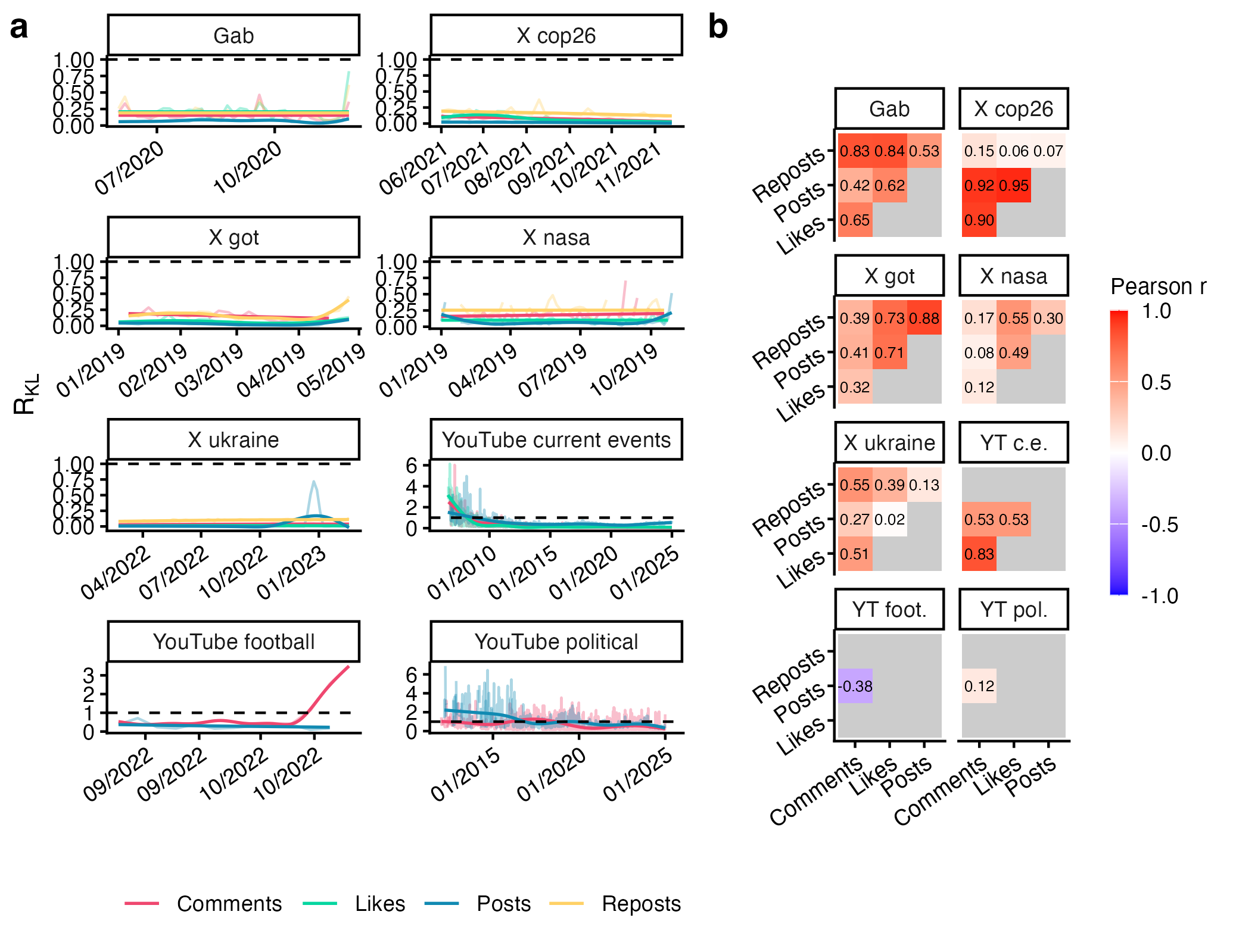}
  \caption{\textbf{Distribution profiles relative to power-law and exponential behavior.} We note how the shape of the distributions is consistently closer to a power-law than to an exponential (\textbf{a}). Values for the statistics being almost always $< 1$ indicate that the divergence between the empirical distribution and a power-law distribution is smaller than the divergence between an exponential and the empirical distribution. Panel (\textbf{b}) displays correlation values between interaction types, divided by platform.}
  \label{fig.divergence_ratio}
\end{figure}

Figure \ref{fig.divergence_ratio}(a) reports the ratio between the divergence from an exponential distribution and the divergence from a power-law distribution, as defined in Equation \eqref{eq:ratio} and introduced in Section \ref{sec:measures}. Recall that a ratio smaller than $1$ indicates that the empirical distribution is better approximated by a power-law model, whereas values larger than $1$ suggest a better approximation by an exponential distribution. Across almost all datasets and interaction types, the observed values remain consistently below $1$. This provides strong evidence that the empirical distributions are systematically closer to a power-law form than to an exponential one. In other words, the heavy-tailed behavior commonly associated with power-law distributions appears to characterize the interaction dynamics of the platforms under study. An additional observation is that there is only limited variation between active and passive interaction measures. Here, active interactions refer to content generated directly by users, such as the number of posts created, while passive interactions capture the engagement accumulated by that content, such as reactions or responses received. Despite describing different aspects of user behavior, both categories display remarkably similar divergence ratios, suggesting that the same underlying statistical structure governs both content production and engagement processes.

Figure \ref{fig.divergence_ratio}(b) further reports the correlation between the divergence ratio time series calculated of different interactions of each dataset. In nearly all cases, correlations are either negligible or positive. The absence of strong negative correlations indicates that shifts toward more power-law-like behavior in one metric are generally not associated with opposite trends in another. Instead, the weak-to-positive correlations suggest that these distributional properties tend to evolve coherently across interaction types. This consistency supports the interpretation that the emergence and persistence of heavy-tailed distributions are not isolated phenomena tied to a single behavioral signal, but rather reflect an organic and platform-wide characteristic of user activity and engagement dynamics. To further investigate the temporal evolution of this measure, Table \ref{tab:mk_klratio} reports the results of the Mann--Kendall trend test applied to the divergence ratio time series. Negative values of Kendall's $\tau$ indicate decreasing trends over time, corresponding to progressively smaller divergence ratios and therefore to distributions becoming increasingly closer to a power-law form. Across most datasets, the estimated $\tau$ statistics are negative, although the strength and statistical significance of the trend vary substantially across platforms and interaction types.

The clearest evidence of decreasing trends is observed in the X COP26 dataset, where all interaction statistics exhibit strongly negative and highly significant coefficients. Similar, though generally weaker, decreasing patterns also emerge in several Gab and YouTube datasets. These results suggest that, in many cases, interaction distributions evolve toward increasingly heavy-tailed configurations over time. However, this behavior is not universal. In the X Ukraine dataset, for example, several passive interaction metrics display significant positive trends, indicating a relative movement away from a power-law approximation. The X NASA dataset instead appears largely stable, with small and statistically insignificant coefficients across all measures. Overall, the Mann--Kendall analysis confirms that while power-law-like behavior is pervasive, its temporal evolution remains dependent on the specific platform and discussion context.

\subsection{Inverse coefficient of variation}
One of the most immediate and interpretable measures of concentration is the inverse coefficient of variation (ICV), defined in Equation \eqref{eq:ICV}, computed on the different interaction statistics. This quantity provides insight into the overall shape and dispersion of the distribution. Since the coefficient compares the mean to the standard deviation, the statistic tends toward zero when the variability of the data becomes large relative to its average value. Conversely, when observations are more concentrated around the mean and the standard deviation is comparatively smaller, the statistic increases. Intuitively, values close to $0$ indicate highly dispersed and heavy-tailed distributions, where a relatively small number of observations account for a large fraction of the activity. Higher values instead correspond to more concentrated distributions, where observations are distributed more uniformly around the mean and extreme values are less dominant.

\begin{figure}[ht]
  \centering
  \includegraphics[width = \linewidth]{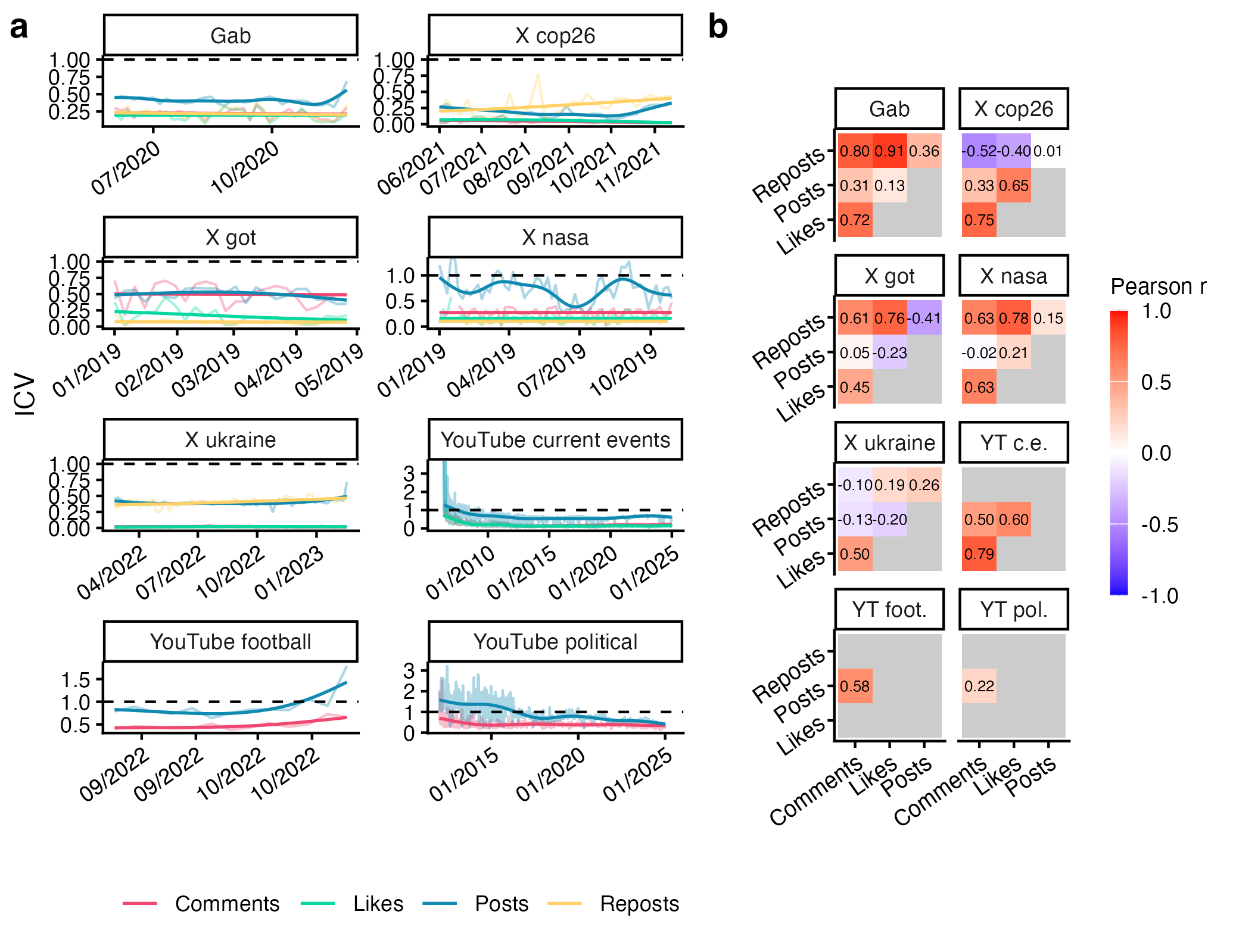}
  \caption{\textbf{Inverse coefficient of variation (ICV) for the different interaction statistics.} The inverse coefficient of variation (\textbf{a}) is almost always less than 1, indicating that, in most cases, the tail of the distribution is long. Panel (\textbf{b}) shows  weak correlations between the measures.}
  \label{fig:meansd}
\end{figure}

The values of this statistic are shown in Figure \ref{fig:meansd}(a). Across most datasets and interaction types, the inverse coefficient of variation remains consistently close to $0$. This suggests that the empirical distributions are generally characterized by substantial dispersion and long tails, in agreement with the evidence previously observed from the divergence-ratio analysis. Values greater than $1$ are only observed sporadically, most notably in the X nasa and YouTube datasets. Interestingly, the statistics that most frequently exceed the threshold of $1$ correspond to active interactions, namely the number of posts created by users within the considered time window. A plausible explanation is that, over relatively short temporal windows, user posting activity may not generate enough extreme events to fully develop a heavy tail. In such situations, the distribution remains comparatively more concentrated, leading to larger values of the inverse coefficient of variation. Passive interactions, such as received engagement, instead appear much more consistently heavy-tailed, likely because attention accumulation processes naturally amplify disparities between highly visible and less visible content.

The correlation subplots in Figure \ref{fig:meansd}(b) again show predominantly non-negative relationships between the different interaction statistics, although the correlations appear weaker than those observed for the divergence-ratio measure. This indicates that while concentration patterns tend to evolve coherently across metrics, the strength of this co-evolution is more moderate when measured through dispersion-based statistics.

To assess whether these concentration levels exhibit systematic temporal trends, Table \ref{tab:mk_icv} reports the results of the Mann--Kendall trend test applied to the inverse coefficient of variation time series. In contrast to the divergence-ratio analysis, most estimated values of Kendall's $\tau$ remain relatively close to zero, indicating the absence of strong monotonic trends over time. This suggests that, although the distributions are consistently highly dispersed, the overall degree of dispersion tends to remain comparatively stable throughout the observation period. Some exceptions nevertheless emerge. In the X Ukraine dataset, several interaction statistics display positive and statistically significant trends, particularly for reposts, suggesting a gradual increase in concentration over time. Conversely, a number of YouTube datasets exhibit negative coefficients, most notably the political dataset for the number of posts, indicating increasing dispersion and heavier tails. The X COP26 dataset presents a more heterogeneous behavior, with opposite trends across different interaction types. Overall, however, the Mann--Kendall analysis confirms that the inverse coefficient of variation is substantially more temporally stable than the divergence-ratio metric.

\subsection{Log-Transformed Gini Index}
\begin{figure}[ht]
  \centering
  \includegraphics[width = \linewidth]{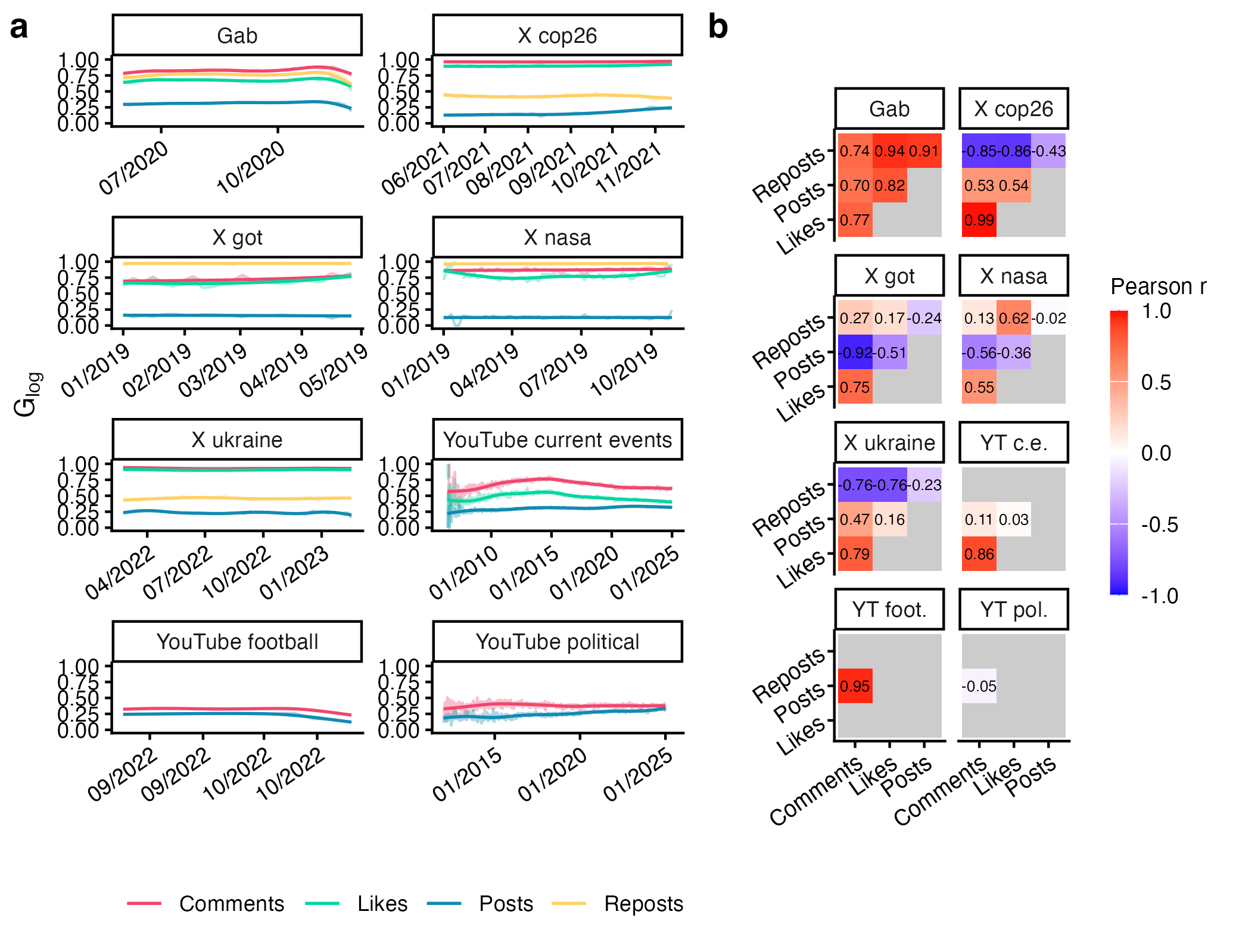}
  \caption{\textbf{Log-transformed Gini index for the different interaction statistics.} The log-transformed Gini index (\textbf{a}) shows high levels of concentration, a phenomenon most apparent in platforms like X or Gab, regardless of the topic of the discussion. Across platforms with multiple datasets, these metrics remain remarkably similar, suggesting underlying platform-specific structural differences. Panel (\textbf{b}) reports correlation values between the measures.}
  \label{fig:loggini}
\end{figure}

Figure \ref{fig:loggini}(a) reports the evolution of the log-transformed Gini index, defined in Equation \eqref{eq:loggini}. The Gini index is a standard measure of inequality and concentration, with larger values corresponding to stronger disparities among observations. In this context, the logarithmic transformation is introduced to attenuate the influence of extremely dominant observations and make differences between datasets more interpretable. Despite this transformation, many of the analyzed distributions still exhibit values of the transformed index close to $G_{log} \approx 1$, indicating that the underlying interaction distributions remain highly concentrated. These results provide further evidence that user activity and engagement are distributed very unevenly in all platforms. A relatively small fraction of users or posts appears to account for a disproportionately large share of the observed interactions, even after reducing the effect of extreme outliers through the logarithmic transformation. This confirms the presence of strong inequality in participation and visibility dynamics across the studied systems.

Two forms of consistency emerge particularly clearly from Figure \ref{fig:loggini}. First, the values of the transformed Gini index are highly stable over time within each dataset. This temporal stability suggests that concentration is not simply the result of isolated events or short-lived fluctuations, but rather a persistent structural property of the interaction dynamics. Second, datasets belonging to the same platform tend to display remarkably similar concentration levels, even when they correspond to different topics or communities. This indicates that the observed degree of inequality is likely influenced by platform-specific mechanisms, such as recommendation systems, visibility algorithms, interaction affordances, or broader patterns of user behavior encouraged by the platform design.

The correlation panel in Figure \ref{fig:loggini}(b) reveals a more heterogeneous pattern than the previous measures. In particular, several interaction statistics in the X datasets exhibit notable negative correlations. However, these negative correlations should be interpreted cautiously. In these datasets, the transformed Gini index varies only minimally over time and remains extremely stable. As a consequence, even very small fluctuations in the measured values may produce comparatively strong correlation coefficients, amplifying the apparent strength of the relationship despite the limited magnitude of the underlying changes.

To further evaluate the temporal dynamics of concentration, Table \ref{tab:mk_loggini} presents the results of the Mann--Kendall trend test applied to the log-transformed Gini coefficient. Consistently with the visual evidence from Figure \ref{fig:loggini}, many of the estimated values of Kendall's $\tau$ remain relatively close to zero, indicating limited monotonic variation over time. This further supports the interpretation that concentration levels are generally stable structural properties of the platforms rather than rapidly evolving characteristics.

At the same time, several datasets exhibit statistically significant positive trends, corresponding to increasing concentration over time. This behavior is especially evident in the X COP26 dataset and in the YouTube political and current events datasets for the number of posts, where the values of $\tau$ are comparatively large. In these cases, the interaction dynamics appear to become progressively more unequal, with activity concentrating increasingly around a smaller subset of users or content. Positive trends are also observed for several passive interaction metrics in the X datasets, particularly likes/favourites and comments/replies. Negative trends, while less common, are also present in some datasets and interaction types. For example, the X GoT and X Ukraine datasets exhibit decreasing concentration for the number of posts, while some YouTube current events engagement metrics display moderate negative coefficients. These results indicate that the temporal evolution of concentration is not uniform across platforms and metrics. Nevertheless, the overall magnitude of the observed trends generally remains moderate compared to the persistently high baseline levels of concentration shown in Figure \ref{fig:loggini}. 

\section{Conclusions}
In this work, we investigated whether inequality in user interactions on social media reflects platform-specific configurations or deeper structural constraints. To this end, we analyzed bipartite user–post interaction networks across multiple platforms and tracked their evolution over time. By examining both active contributions (posts) and passive engagement (likes and comments), we aimed to capture the broader architecture of online participation. We quantified inequality using a KL-divergence–based model comparison, an inverse coefficient of variation, and a log-transformed Gini index. All metrics were computed over time, allowing us to assess whether and how attention distribution patterns evolve across diverse digital environments.

Our results show that interaction inequality is not only widespread but remarkably persistent. Across all platforms, and regardless of differences in scale, governance, or topical focus, a small fraction of users consistently accumulates the majority of attention. This pattern holds for all forms of engagement and exhibits a stable cross-platform ordering over time. Passive interactions tend to be even more concentrated than active ones. These findings point to the conclusion that inequality in digital interaction is not an incidental byproduct of algorithms, content dynamics, or user preferences. Rather, it emerges as a structural outcome of systems built around constrained attention and platform-mediated visibility. Under these conditions, engagement becomes subject to compression: attention gravitates toward a few visible actors, while the majority remain marginal. The recurrence of this pattern across contexts suggests that inequality acts as a systemic attractor—resistant to surface-level interventions.

Understanding these regularities is essential for rethinking the promises and limits of digital participation. Platforms are often designed with the intention of enabling open access, inclusive visibility, and equal opportunity for engagement.  In this context, visibility and attention are scarse and unevenly distributed, but follow stable trajectories over time. Our analysis therefore shows that these policy goals are bounded by the very architecture of digital interactions.

\backmatter








\section*{Declarations}

\subsection*{Availability of data and materials}
\begin{itemize}
    \item \textbf{Gab}, used in prior work: see \cite{dimartino2025}
    \item \textbf{X COP26}, used in prior work: see \cite{falkenberg2022growing}
    \item \textbf{X Ukraine}, used in prior work: see \cite{Loru2025agenda}
    \item \textbf{YouTube} datasets (football, current events and political) are available on \url{https://osf.io/3824q/}
    \item \textbf{X NASA} and \textbf{X got}, used in prior work: see \cite{avalle2024persistent} 
\end{itemize}

\subsection*{Competing interests}
The authors declare no competing interests.
\subsection*{Funding}
Funded by the European Union - Next Generation EU, Mission 4 Component 1 CUP B53C23002460006.
\subsection*{Authors' contributions}
G.P. designed and conducted the experiments; G.P., E.DM., E.L., and S.Z. analyzed the results; N.DM. and M.C. supervised the project; all authors wrote the manuscript and provided critical feedback.


\begin{thebibliography}{39}
\ifx \bisbn   \undefined \def \bisbn  #1{ISBN #1}\fi
\ifx \binits  \undefined \def \binits#1{#1}\fi
\ifx \bauthor  \undefined \def \bauthor#1{#1}\fi
\ifx \batitle  \undefined \def \batitle#1{#1}\fi
\ifx \bjtitle  \undefined \def \bjtitle#1{#1}\fi
\ifx \bvolume  \undefined \def \bvolume#1{\textbf{#1}}\fi
\ifx \byear  \undefined \def \byear#1{#1}\fi
\ifx \bissue  \undefined \def \bissue#1{#1}\fi
\ifx \bfpage  \undefined \def \bfpage#1{#1}\fi
\ifx \blpage  \undefined \def \blpage #1{#1}\fi
\ifx \burl  \undefined \def \burl#1{\textsf{#1}}\fi
\ifx \doiurl  \undefined \def \doiurl#1{\url{https://doi.org/#1}}\fi
\ifx \betal  \undefined \def \betal{\textit{et al.}}\fi
\ifx \binstitute  \undefined \def \binstitute#1{#1}\fi
\ifx \binstitutionaled  \undefined \def \binstitutionaled#1{#1}\fi
\ifx \bctitle  \undefined \def \bctitle#1{#1}\fi
\ifx \beditor  \undefined \def \beditor#1{#1}\fi
\ifx \bpublisher  \undefined \def \bpublisher#1{#1}\fi
\ifx \bbtitle  \undefined \def \bbtitle#1{#1}\fi
\ifx \bedition  \undefined \def \bedition#1{#1}\fi
\ifx \bseriesno  \undefined \def \bseriesno#1{#1}\fi
\ifx \blocation  \undefined \def \blocation#1{#1}\fi
\ifx \bsertitle  \undefined \def \bsertitle#1{#1}\fi
\ifx \bsnm \undefined \def \bsnm#1{#1}\fi
\ifx \bsuffix \undefined \def \bsuffix#1{#1}\fi
\ifx \bparticle \undefined \def \bparticle#1{#1}\fi
\ifx \barticle \undefined \def \barticle#1{#1}\fi
\bibcommenthead
\ifx \bconfdate \undefined \def \bconfdate #1{#1}\fi
\ifx \botherref \undefined \def \botherref #1{#1}\fi
\ifx \url \undefined \def \url#1{\textsf{#1}}\fi
\ifx \bchapter \undefined \def \bchapter#1{#1}\fi
\ifx \bbook \undefined \def \bbook#1{#1}\fi
\ifx \bcomment \undefined \def \bcomment#1{#1}\fi
\ifx \oauthor \undefined \def \oauthor#1{#1}\fi
\ifx \citeauthoryear \undefined \def \citeauthoryear#1{#1}\fi
\ifx \endbibitem  \undefined \def \endbibitem {}\fi
\ifx \bconflocation  \undefined \def \bconflocation#1{#1}\fi
\ifx \arxivurl  \undefined \def \arxivurl#1{\textsf{#1}}\fi
\csname PreBibitemsHook\endcsname

\bibitem[\protect\citeauthoryear{Di~Marco et~al.}{2026}]{DiMarco2026patterns}
\begin{barticle}
\bauthor{\bsnm{Di~Marco}, \binits{N.}},
\bauthor{\bsnm{Bonetti}, \binits{A.}},
\bauthor{\bsnm{Di~Martino}, \binits{E.}},
\bauthor{\bsnm{Loru}, \binits{E.}},
\bauthor{\bsnm{Nudo}, \binits{J.}},
\bauthor{\bsnm{Pandolfo}, \binits{M.E.}},
\bauthor{\bsnm{Pecile}, \binits{G.}},
\bauthor{\bsnm{Sangiorgio}, \binits{E.}},
\bauthor{\bsnm{Scalco}, \binits{I.}},
\bauthor{\bsnm{Zollo}, \binits{S.}},
\bauthor{\bsnm{Cinelli}, \binits{M.}},
\bauthor{\bsnm{Zollo}, \binits{F.}},
\bauthor{\bsnm{Quattrociocchi}, \binits{W.}}:
\batitle{Patterns, models, and challenges in online social media: A survey}.
\bjtitle{ACM Transactions on the Web}
\bvolume{20}(\bissue{2}),
\bfpage{1}--\blpage{34}
(\byear{2026})
\doiurl{10.1145/3796548}
\end{barticle}
\endbibitem

\bibitem[\protect\citeauthoryear{Diaz~Ruiz and Nilsson}{2023}]{diaz2023disinformation}
\begin{barticle}
\bauthor{\bsnm{Diaz~Ruiz}, \binits{C.}},
\bauthor{\bsnm{Nilsson}, \binits{T.}}:
\batitle{Disinformation and echo chambers: how disinformation circulates on social media through identity-driven controversies}.
\bjtitle{Journal of public policy \& marketing}
\bvolume{42}(\bissue{1}),
\bfpage{18}--\blpage{35}
(\byear{2023})
\end{barticle}
\endbibitem

\bibitem[\protect\citeauthoryear{Budak et~al.}{2024}]{Budak2024_misinfo}
\begin{barticle}
\bauthor{\bsnm{Budak}, \binits{C.}},
\bauthor{\bsnm{Nyhan}, \binits{B.}},
\bauthor{\bsnm{Rothschild}, \binits{D.M.}},
\bauthor{\bsnm{Thorson}, \binits{E.}},
\bauthor{\bsnm{Watts}, \binits{D.J.}}:
\batitle{Misunderstanding the harms of online misinformation}.
\bjtitle{Nature}
\bvolume{630}(\bissue{8015}),
\bfpage{45}--\blpage{53}
(\byear{2024})
\doiurl{10.1038/s41586-024-07417-w}
\end{barticle}
\endbibitem

\bibitem[\protect\citeauthoryear{Avalle et~al.}{2024}]{avalle2024persistent}
\begin{barticle}
\bauthor{\bsnm{Avalle}, \binits{M.}},
\bauthor{\bsnm{Di~Marco}, \binits{N.}},
\bauthor{\bsnm{Etta}, \binits{G.}},
\bauthor{\bsnm{Sangiorgio}, \binits{E.}},
\bauthor{\bsnm{Alipour}, \binits{S.}},
\bauthor{\bsnm{Bonetti}, \binits{A.}},
\bauthor{\bsnm{Alvisi}, \binits{L.}},
\bauthor{\bsnm{Scala}, \binits{A.}},
\bauthor{\bsnm{Baronchelli}, \binits{A.}},
\bauthor{\bsnm{Cinelli}, \binits{M.}}, \betal:
\batitle{Persistent interaction patterns across social media platforms and over time}.
\bjtitle{Nature}
\bvolume{628}(\bissue{8008}),
\bfpage{582}--\blpage{589}
(\byear{2024})
\end{barticle}
\endbibitem

\bibitem[\protect\citeauthoryear{Tahmasbi et~al.}{2021}]{tahmasbi2021go}
\begin{bchapter}
\bauthor{\bsnm{Tahmasbi}, \binits{F.}},
\bauthor{\bsnm{Schild}, \binits{L.}},
\bauthor{\bsnm{Ling}, \binits{C.}},
\bauthor{\bsnm{Blackburn}, \binits{J.}},
\bauthor{\bsnm{Stringhini}, \binits{G.}},
\bauthor{\bsnm{Zhang}, \binits{Y.}},
\bauthor{\bsnm{Zannettou}, \binits{S.}}:
\bctitle{“go eat a bat, chang!”: On the emergence of sinophobic behavior on web communities in the face of covid-19}.
In: \bbtitle{Proceedings of the Web Conference 2021},
pp. \bfpage{1122}--\blpage{1133}
(\byear{2021})
\end{bchapter}
\endbibitem

\bibitem[\protect\citeauthoryear{Loru et~al.}{2024}]{lorutox}
\begin{barticle}
\bauthor{\bsnm{Loru}, \binits{E.}},
\bauthor{\bsnm{Cinelli}, \binits{M.}},
\bauthor{\bsnm{Tesconi}, \binits{M.}},
\bauthor{\bsnm{Quattrociocchi}, \binits{W.}}:
\batitle{The influence of coordinated behavior on toxicity}.
\bjtitle{Online Social Networks and Media}
\bvolume{43-44},
\bfpage{100289}
(\byear{2024})
\doiurl{10.1016/j.osnem.2024.100289}
\end{barticle}
\endbibitem

\bibitem[\protect\citeauthoryear{Falkenberg et~al.}{2024}]{falkenberg24patterns}
\begin{botherref}
\oauthor{\bsnm{Falkenberg}, \binits{M.}},
\oauthor{\bsnm{Zollo}, \binits{F.}},
\oauthor{\bsnm{Quattrociocchi}, \binits{W.}},
\oauthor{\bsnm{Pfeffer}, \binits{J.}},
\oauthor{\bsnm{Baronchelli}, \binits{A.}}:
Patterns of partisan toxicity and engagement reveal the common structure of online political communication across countries.
Nature Communications
\textbf{15}
(2024)
\doiurl{10.1038/s41467-024-53868-0}
\end{botherref}
\endbibitem

\bibitem[\protect\citeauthoryear{Bail et~al.}{2018}]{bail2024exposure}
\begin{barticle}
\bauthor{\bsnm{Bail}, \binits{C.A.}},
\bauthor{\bsnm{Argyle}, \binits{L.P.}},
\bauthor{\bsnm{Brown}, \binits{T.W.}},
\bauthor{\bsnm{Bumpus}, \binits{J.P.}},
\bauthor{\bsnm{Chen}, \binits{H.}},
\bauthor{\bsnm{Hunzaker}, \binits{M.B.F.}},
\bauthor{\bsnm{Lee}, \binits{J.}},
\bauthor{\bsnm{Mann}, \binits{M.}},
\bauthor{\bsnm{Merhout}, \binits{F.}},
\bauthor{\bsnm{Volfovsky}, \binits{A.}}:
\batitle{Exposure to opposing views on social media can increase political polarization}.
\bjtitle{Proceedings of the National Academy of Sciences}
\bvolume{115}(\bissue{37}),
\bfpage{9216}--\blpage{9221}
(\byear{2018})
\doiurl{10.1073/pnas.1804840115}
{\href{https://arxiv.org/abs/https://www.pnas.org/doi/pdf/10.1073/pnas.1804840115}{{https://www.pnas.org/doi/pdf/10.1073/pnas.1804840115}}}
\end{barticle}
\endbibitem

\bibitem[\protect\citeauthoryear{Kubin and Von~Sikorski}{2021}]{kubin2021role}
\begin{barticle}
\bauthor{\bsnm{Kubin}, \binits{E.}},
\bauthor{\bsnm{Von~Sikorski}, \binits{C.}}:
\batitle{The role of (social) media in political polarization: a systematic review}.
\bjtitle{Annals of the International Communication Association}
\bvolume{45}(\bissue{3}),
\bfpage{188}--\blpage{206}
(\byear{2021})
\end{barticle}
\endbibitem

\bibitem[\protect\citeauthoryear{Pecile et~al.}{2025}]{pecile2025}
\begin{bchapter}
\bauthor{\bsnm{Pecile}, \binits{G.}},
\bauthor{\bsnm{Di~Marco}, \binits{N.}},
\bauthor{\bsnm{Cinelli}, \binits{M.}},
\bauthor{\bsnm{Quattrociocchi}, \binits{W.}}:
\bctitle{Decoding political polarization in social media interactions}.
In: \beditor{\bsnm{Cherifi}, \binits{H.}},
\beditor{\bsnm{Donduran}, \binits{M.}},
\beditor{\bsnm{Rocha}, \binits{L.M.}},
\beditor{\bsnm{Cherifi}, \binits{C.}},
\beditor{\bsnm{Varol}, \binits{O.}} (eds.)
\bbtitle{Complex Networks {\&} Their Applications XIII},
pp. \bfpage{282}--\blpage{293}.
\bpublisher{Springer},
\blocation{Cham}
(\byear{2025})
\end{bchapter}
\endbibitem

\bibitem[\protect\citeauthoryear{Falkenberg et~al.}{2023}]{falkenberg23affective}
\begin{botherref}
\oauthor{\bsnm{Falkenberg}, \binits{M.}},
\oauthor{\bsnm{Zollo}, \binits{F.}},
\oauthor{\bsnm{Quattrociocchi}, \binits{W.}},
\oauthor{\bsnm{Pfeffer}, \binits{J.}},
\oauthor{\bsnm{Baronchelli}, \binits{A.}}:
Affective and interactional polarization align across countries
(2023)
\doiurl{10.31234/osf.io/ngsb6_v1}
\end{botherref}
\endbibitem

\bibitem[\protect\citeauthoryear{Loru et~al.}{2025}]{Loru2025agenda}
\begin{botherref}
\oauthor{\bsnm{Loru}, \binits{E.}},
\oauthor{\bsnm{Galeazzi}, \binits{A.}},
\oauthor{\bsnm{Bonetti}, \binits{A.}},
\oauthor{\bsnm{Sangiorgio}, \binits{E.}},
\oauthor{\bsnm{Di~Marco}, \binits{N.}},
\oauthor{\bsnm{Cinelli}, \binits{M.}},
\oauthor{\bsnm{Falkenberg}, \binits{M.}},
\oauthor{\bsnm{Baronchelli}, \binits{A.}},
\oauthor{\bsnm{Quattrociocchi}, \binits{W.}}:
Ideology and polarization set the agenda on social media.
Scientific Reports
\textbf{15}(1)
(2025)
\doiurl{10.1038/s41598-025-19776-z}
\end{botherref}
\endbibitem

\bibitem[\protect\citeauthoryear{for Countering Digital~Hate}{2022}]{disinfo12}
\begin{botherref}
\oauthor{\bsnm{Countering Digital~Hate}, \binits{C.}}:
THE DISINFORMATION DOZEN - Why platforms must act on twelve leading online anti-vaxxers
(2022).
\url{https://counterhate.com/wp-content/uploads/2022/05/210324-The-Disinformation-Dozen.pdf}
\end{botherref}
\endbibitem

\bibitem[\protect\citeauthoryear{Scalco et~al.}{2026}]{scalco2026detect}
\begin{botherref}
\oauthor{\bsnm{Scalco}, \binits{I.}},
\oauthor{\bsnm{Gesualdo}, \binits{F.}},
\oauthor{\bsnm{Cerqueti}, \binits{R.}},
\oauthor{\bsnm{Cinelli}, \binits{M.}}:
How to detect information voids using longitudinal data from social media and web searches.
arXiv preprint arXiv:2602.15476
(2026)
\end{botherref}
\endbibitem

\bibitem[\protect\citeauthoryear{Adamic and Huberman}{2000}]{adamic2000power}
\begin{barticle}
\bauthor{\bsnm{Adamic}, \binits{L.A.}},
\bauthor{\bsnm{Huberman}, \binits{B.A.}}:
\batitle{Power-law distribution of the world wide web}.
\bjtitle{science}
\bvolume{287}(\bissue{5461}),
\bfpage{2115}--\blpage{2115}
(\byear{2000})
\end{barticle}
\endbibitem

\bibitem[\protect\citeauthoryear{Muchnik et~al.}{2013}]{muchnik2013origins}
\begin{barticle}
\bauthor{\bsnm{Muchnik}, \binits{L.}},
\bauthor{\bsnm{Pei}, \binits{S.}},
\bauthor{\bsnm{Parra}, \binits{L.C.}},
\bauthor{\bsnm{Reis}, \binits{S.D.}},
\bauthor{\bsnm{Andrade~Jr}, \binits{J.S.}},
\bauthor{\bsnm{Havlin}, \binits{S.}},
\bauthor{\bsnm{Makse}, \binits{H.A.}}:
\batitle{Origins of power-law degree distribution in the heterogeneity of human activity in social networks}.
\bjtitle{Scientific reports}
\bvolume{3}(\bissue{1}),
\bfpage{1783}
(\byear{2013})
\end{barticle}
\endbibitem

\bibitem[\protect\citeauthoryear{Asur et~al.}{2011}]{asur2011trends}
\begin{bchapter}
\bauthor{\bsnm{Asur}, \binits{S.}},
\bauthor{\bsnm{Huberman}, \binits{B.A.}},
\bauthor{\bsnm{Szabo}, \binits{G.}},
\bauthor{\bsnm{Wang}, \binits{C.}}:
\bctitle{Trends in social media: Persistence and decay}.
In: \bbtitle{Proceedings of the International AAAI Conference on Web and Social Media},
vol. \bseriesno{5},
pp. \bfpage{434}--\blpage{437}
(\byear{2011})
\end{bchapter}
\endbibitem

\bibitem[\protect\citeauthoryear{Ross and Jones}{2015}]{ross2015understanding}
\begin{barticle}
\bauthor{\bsnm{Ross}, \binits{G.J.}},
\bauthor{\bsnm{Jones}, \binits{T.}}:
\batitle{Understanding the heavy-tailed dynamics in human behavior}.
\bjtitle{Physical Review E}
\bvolume{91}(\bissue{6}),
\bfpage{062809}
(\byear{2015})
\end{barticle}
\endbibitem

\bibitem[\protect\citeauthoryear{Mathews et~al.}{2017}]{mathews2017nature}
\begin{bchapter}
\bauthor{\bsnm{Mathews}, \binits{P.}},
\bauthor{\bsnm{Mitchell}, \binits{L.}},
\bauthor{\bsnm{Nguyen}, \binits{G.}},
\bauthor{\bsnm{Bean}, \binits{N.}}:
\bctitle{The nature and origin of heavy tails in retweet activity}.
In: \bbtitle{Proceedings of the 26th International Conference on World Wide Web Companion},
pp. \bfpage{1493}--\blpage{1498}
(\byear{2017})
\end{bchapter}
\endbibitem

\bibitem[\protect\citeauthoryear{Di~Marco et~al.}{2024}]{di2024patterns}
\begin{barticle}
\bauthor{\bsnm{Di~Marco}, \binits{N.}},
\bauthor{\bsnm{Loru}, \binits{E.}},
\bauthor{\bsnm{Bonetti}, \binits{A.}},
\bauthor{\bsnm{Serra}, \binits{A.O.G.}},
\bauthor{\bsnm{Cinelli}, \binits{M.}},
\bauthor{\bsnm{Quattrociocchi}, \binits{W.}}:
\batitle{Patterns of linguistic simplification on social media platforms over time}.
\bjtitle{Proceedings of the National Academy of Sciences}
\bvolume{121}(\bissue{50}),
\bfpage{2412105121}
(\byear{2024})
\end{barticle}
\endbibitem

\bibitem[\protect\citeauthoryear{Robins et~al.}{2009}]{robins2009closure}
\begin{barticle}
\bauthor{\bsnm{Robins}, \binits{G.}},
\bauthor{\bsnm{Pattison}, \binits{P.}},
\bauthor{\bsnm{Wang}, \binits{P.}}:
\batitle{Closure, connectivity and degree distributions: Exponential random graph (p*) models for directed social networks}.
\bjtitle{Social networks}
\bvolume{31}(\bissue{2}),
\bfpage{105}--\blpage{117}
(\byear{2009})
\end{barticle}
\endbibitem

\bibitem[\protect\citeauthoryear{Ahn et~al.}{2007}]{ahn2007analysis}
\begin{bchapter}
\bauthor{\bsnm{Ahn}, \binits{Y.-Y.}},
\bauthor{\bsnm{Han}, \binits{S.}},
\bauthor{\bsnm{Kwak}, \binits{H.}},
\bauthor{\bsnm{Moon}, \binits{S.}},
\bauthor{\bsnm{Jeong}, \binits{H.}}:
\bctitle{Analysis of topological characteristics of huge online social networking services}.
In: \bbtitle{Proceedings of the 16th International Conference on World Wide Web},
pp. \bfpage{835}--\blpage{844}
(\byear{2007})
\end{bchapter}
\endbibitem

\bibitem[\protect\citeauthoryear{Agichtein et~al.}{2008}]{agichtein2008finding}
\begin{bchapter}
\bauthor{\bsnm{Agichtein}, \binits{E.}},
\bauthor{\bsnm{Castillo}, \binits{C.}},
\bauthor{\bsnm{Donato}, \binits{D.}},
\bauthor{\bsnm{Gionis}, \binits{A.}},
\bauthor{\bsnm{Mishne}, \binits{G.}}:
\bctitle{Finding high-quality content in social media}.
In: \bbtitle{Proceedings of the 2008 International Conference on Web Search and Data Mining},
pp. \bfpage{183}--\blpage{194}
(\byear{2008})
\end{bchapter}
\endbibitem

\bibitem[\protect\citeauthoryear{Zafarani et~al.}{2014}]{zafarani2014social}
\begin{bbook}
\bauthor{\bsnm{Zafarani}, \binits{R.}},
\bauthor{\bsnm{Abbasi}, \binits{M.A.}},
\bauthor{\bsnm{Liu}, \binits{H.}}:
\bbtitle{Social Media Mining: an Introduction}.
\bpublisher{Cambridge University Press}, \blocation{???}
(\byear{2014})
\end{bbook}
\endbibitem

\bibitem[\protect\citeauthoryear{Cinelli et~al.}{2020}]{cinelli2020selective}
\begin{barticle}
\bauthor{\bsnm{Cinelli}, \binits{M.}},
\bauthor{\bsnm{Brugnoli}, \binits{E.}},
\bauthor{\bsnm{Schmidt}, \binits{A.L.}},
\bauthor{\bsnm{Zollo}, \binits{F.}},
\bauthor{\bsnm{Quattrociocchi}, \binits{W.}},
\bauthor{\bsnm{Scala}, \binits{A.}}:
\batitle{Selective exposure shapes the facebook news diet}.
\bjtitle{PloS one}
\bvolume{15}(\bissue{3}),
\bfpage{0229129}
(\byear{2020})
\end{barticle}
\endbibitem

\bibitem[\protect\citeauthoryear{Sangiorgio et~al.}{2025}]{sangiorgio2025evaluating}
\begin{barticle}
\bauthor{\bsnm{Sangiorgio}, \binits{E.}},
\bauthor{\bsnm{Di~Marco}, \binits{N.}},
\bauthor{\bsnm{Etta}, \binits{G.}},
\bauthor{\bsnm{Cinelli}, \binits{M.}},
\bauthor{\bsnm{Cerqueti}, \binits{R.}},
\bauthor{\bsnm{Quattrociocchi}, \binits{W.}}:
\batitle{Evaluating the effect of viral posts on social media engagement}.
\bjtitle{Scientific Reports}
\bvolume{15}(\bissue{1}),
\bfpage{639}
(\byear{2025})
\end{barticle}
\endbibitem

\bibitem[\protect\citeauthoryear{Di~Marco et~al.}{2024}]{DiMarco2024_volatility}
\begin{barticle}
\bauthor{\bsnm{Di~Marco}, \binits{N.}},
\bauthor{\bsnm{Cinelli}, \binits{M.}},
\bauthor{\bsnm{Alipour}, \binits{S.}},
\bauthor{\bsnm{Quattrociocchi}, \binits{W.}}:
\batitle{Users volatility on reddit and voat}.
\bjtitle{IEEE Transactions on Computational Social Systems}
\bvolume{11}(\bissue{5}),
\bfpage{5871}--\blpage{5879}
(\byear{2024})
\doiurl{10.1109/tcss.2024.3379318}
\end{barticle}
\endbibitem

\bibitem[\protect\citeauthoryear{Zollo et~al.}{2026}]{zollo2026examining}
\begin{barticle}
\bauthor{\bsnm{Zollo}, \binits{S.}},
\bauthor{\bsnm{Sangiorgio}, \binits{E.}},
\bauthor{\bsnm{Etta}, \binits{G.}},
\bauthor{\bsnm{{Di Marco}}, \binits{N.}},
\bauthor{\bsnm{Cinelli}, \binits{M.}},
\bauthor{\bsnm{Cerqueti}, \binits{R.}},
\bauthor{\bsnm{Quattrociocchi}, \binits{W.}}:
\batitle{Examining the relationship between content length and engagement with news outlets on multiple social media platforms}.
\bjtitle{Technological Forecasting and Social Change}
\bvolume{228},
\bfpage{124668}
(\byear{2026})
\doiurl{10.1016/j.techfore.2026.124668}
\end{barticle}
\endbibitem

\bibitem[\protect\citeauthoryear{Machado et~al.}{2025}]{machado2025super}
\begin{botherref}
\oauthor{\bsnm{Machado}, \binits{G.}},
\oauthor{\bsnm{Pacheco}, \binits{D.}},
\oauthor{\bsnm{Menezes}, \binits{R.}},
\oauthor{\bsnm{Baxter}, \binits{G.}}:
Super-linear growth and rising inequality in online social communities: Insights from reddit.
arXiv preprint arXiv:2503.02661
(2025)
\end{botherref}
\endbibitem

\bibitem[\protect\citeauthoryear{Panek et~al.}{2018}]{Panek2018}
\begin{barticle}
\bauthor{\bsnm{Panek}, \binits{E.}},
\bauthor{\bsnm{Hollenbach}, \binits{C.}},
\bauthor{\bsnm{Yang}, \binits{J.}},
\bauthor{\bsnm{Rhodes}, \binits{T.}}:
\batitle{The effects of group size and time on the formation of online communities: Evidence from reddit}.
\bjtitle{Social Media + Society}
\bvolume{4},
\bfpage{205630511881590}
(\byear{2018})
\doiurl{10.1177/2056305118815908}
\end{barticle}
\endbibitem

\bibitem[\protect\citeauthoryear{Singer et~al.}{2014}]{singer14}
\begin{bchapter}
\bauthor{\bsnm{Singer}, \binits{P.}},
\bauthor{\bsnm{Fl\"{o}ck}, \binits{F.}},
\bauthor{\bsnm{Meinhart}, \binits{C.}},
\bauthor{\bsnm{Zeitfogel}, \binits{E.}},
\bauthor{\bsnm{Strohmaier}, \binits{M.}}:
\bctitle{Evolution of reddit: from the front page of the internet to a self-referential community?}
In: \bbtitle{Proceedings of the 23rd International Conference on World Wide Web}.
\bsertitle{WWW '14 Companion},
pp. \bfpage{517}--\blpage{522}.
\bpublisher{Association for Computing Machinery},
\blocation{New York, NY, USA}
(\byear{2014}).
\doiurl{10.1145/2567948.2576943} .
\burl{https://doi.org/10.1145/2567948.2576943}
\end{bchapter}
\endbibitem

\bibitem[\protect\citeauthoryear{Orellana-Rodriguez and Keane}{2018}]{ORELLANARODRIGUEZ201874}
\begin{barticle}
\bauthor{\bsnm{Orellana-Rodriguez}, \binits{C.}},
\bauthor{\bsnm{Keane}, \binits{M.T.}}:
\batitle{Attention to news and its dissemination on twitter: A survey}.
\bjtitle{Computer Science Review}
\bvolume{29},
\bfpage{74}--\blpage{94}
(\byear{2018})
\doiurl{10.1016/j.cosrev.2018.07.001}
\end{barticle}
\endbibitem

\bibitem[\protect\citeauthoryear{Zhu and Lerman}{2016}]{zhu2016attention}
\begin{botherref}
\oauthor{\bsnm{Zhu}, \binits{L.}},
\oauthor{\bsnm{Lerman}, \binits{K.}}:
Attention inequality in social media.
arXiv preprint arXiv:1601.07200
(2016)
\end{botherref}
\endbibitem

\bibitem[\protect\citeauthoryear{Bagdouri}{2021}]{Bagdouri_2021}
\begin{barticle}
\bauthor{\bsnm{Bagdouri}, \binits{M.}}:
\batitle{Journalists and twitter: A multidimensional quantitative description of usage patterns}.
\bjtitle{Proceedings of the International AAAI Conference on Web and Social Media}
\bvolume{10}(\bissue{1}),
\bfpage{22}--\blpage{31}
(\byear{2021})
\doiurl{10.1609/icwsm.v10i1.14742}
\end{barticle}
\endbibitem

\bibitem[\protect\citeauthoryear{Orellana-Rodriguez et~al.}{2017}]{Rodriguez17}
\begin{barticle}
\bauthor{\bsnm{Orellana-Rodriguez}, \binits{C.}},
\bauthor{\bsnm{Greene}, \binits{D.}},
\bauthor{\bsnm{Keane}, \binits{M.}}:
\batitle{Spreading one’s tweets: How can journalists gain attention for their tweeted news?}
\bjtitle{Journal of Web Science}
\bvolume{3},
\bfpage{16}--\blpage{31}
(\byear{2017})
\doiurl{10.1561/106.00000009}
\end{barticle}
\endbibitem

\bibitem[\protect\citeauthoryear{Falkenberg et~al.}{2022}]{falkenberg2022growing}
\begin{barticle}
\bauthor{\bsnm{Falkenberg}, \binits{M.}},
\bauthor{\bsnm{Galeazzi}, \binits{A.}},
\bauthor{\bsnm{Torricelli}, \binits{M.}},
\bauthor{\bsnm{Di~Marco}, \binits{N.}},
\bauthor{\bsnm{Larosa}, \binits{F.}},
\bauthor{\bsnm{Sas}, \binits{M.}},
\bauthor{\bsnm{Mekacher}, \binits{A.}},
\bauthor{\bsnm{Pearce}, \binits{W.}},
\bauthor{\bsnm{Zollo}, \binits{F.}},
\bauthor{\bsnm{Quattrociocchi}, \binits{W.}}, \betal:
\batitle{Growing polarization around climate change on social media}.
\bjtitle{Nature Climate Change}
\bvolume{12}(\bissue{12}),
\bfpage{1114}--\blpage{1121}
(\byear{2022})
\end{barticle}
\endbibitem

\bibitem[\protect\citeauthoryear{Chen and Ferrara}{2023}]{Chen_Ferrara_2023}
\begin{barticle}
\bauthor{\bsnm{Chen}, \binits{E.}},
\bauthor{\bsnm{Ferrara}, \binits{E.}}:
\batitle{Tweets in time of conflict: A public dataset tracking the twitter discourse on the war between ukraine and russia}.
\bjtitle{Proceedings of the International AAAI Conference on Web and Social Media}
\bvolume{17}(\bissue{1}),
\bfpage{1006}--\blpage{1013}
(\byear{2023})
\doiurl{10.1609/icwsm.v17i1.22208}
\end{barticle}
\endbibitem

\bibitem[\protect\citeauthoryear{Di~Martino et~al.}{2025}]{dimartino2025}
\begin{botherref}
\oauthor{\bsnm{Di~Martino}, \binits{E.}},
\oauthor{\bsnm{Galeazzi}, \binits{A.}},
\oauthor{\bsnm{Starnini}, \binits{M.}},
\oauthor{\bsnm{Quattrociocchi}, \binits{W.}},
\oauthor{\bsnm{Cinelli}, \binits{M.}}:
Ideological fragmentation of the social media ecosystem: From echo chambers to echo platforms.
PNAS Nexus,
262
(2025)
\doiurl{10.1093/pnasnexus/pgaf262}
{\href{https://arxiv.org/abs/https://academic.oup.com/pnasnexus/advance-article-pdf/doi/10.1093/pnasnexus/pgaf262/64030149/pgaf262.pdf}{{https://academic.oup.com/pnasnexus/advance-article-pdf/doi/10.1093/pnasnexus/pgaf262/64030149/pgaf262.pdf}}}
\end{botherref}
\endbibitem

\bibitem[\protect\citeauthoryear{Clauset et~al.}{2009}]{clauset.2009}
\begin{barticle}
\bauthor{\bsnm{Clauset}, \binits{A.}},
\bauthor{\bsnm{Shalizi}, \binits{C.R.}},
\bauthor{\bsnm{Newman}, \binits{M.E.J.}}:
\batitle{Power-law distributions in empirical data}.
\bjtitle{SIAM Review}
\bvolume{51}(\bissue{4}),
\bfpage{661}--\blpage{703}
(\byear{2009})
\doiurl{10.1137/070710111}
{\href{https://arxiv.org/abs/https://doi.org/10.1137/070710111}{{https://doi.org/10.1137/070710111}}}
\end{barticle}
\endbibitem

\end{thebibliography}

\section*{Supporting information}
\renewcommand{\thefigure}{S\arabic{figure}}
\setcounter{figure}{0}
\renewcommand{\thetable}{S\arabic{table}}
\setcounter{table}{0}
\begin{table}[ht]
\caption{\textbf{Mann--Kendall trend statistics for the divergence ratio.} Negative values of Kendall's $\tau$ indicate decreasing divergence ratios over time, corresponding to empirical distributions becoming progressively closer to a power-law approximation relative to an exponential one. Most datasets exhibit negative trends, although the strength and statistical significance of the effect vary across platforms and interaction types.}
\label{tab:mk_klratio}
\centering
\begin{tabular}{llrr}
\toprule
Dataset & Variable & $\tau$ & p-value (sl)\\
\midrule
Gab & Number of posts & -0.024 & 0.844\\
Gab & Comments/replies & -0.273 & 0.018\\
Gab & Retweets/reblogs & -0.243 & 0.035\\
Gab & Likes/favourites & -0.153 & 0.187\\
\addlinespace
X got & Number of posts & -0.232 & 0.118\\
X got & Likes/favourites & -0.290 & 0.050\\
X got & Retweets/reblogs & -0.170 & 0.267\\
X got & Comments/replies & -0.357 & 0.036\\
\addlinespace
X nasa & Number of posts & 0.124 & 0.171\\
X nasa & Likes/favourites & 0.032 & 0.729\\
X nasa & Retweets/reblogs & 0.107 & 0.304\\
X nasa & Comments/replies & 0.057 & 0.576\\
\addlinespace
X cop26 & Number of posts & -0.482 & 0.000\\
X cop26 & Likes/favourites & -0.479 & 0.000\\
X cop26 & Retweets/reblogs & -0.449 & 0.000\\
X cop26 & Comments/replies & -0.580 & 0.000\\
\addlinespace
X ukraine & Number of posts & -0.081 & 0.310\\
X ukraine & Likes/favourites & 0.166 & 0.038\\
X ukraine & Retweets/reblogs & 0.435 & 0.000\\
X ukraine & Comments/replies & 0.540 & 0.000\\
\addlinespace
YouTube political & Number of posts & -0.352 & 0.000\\
YouTube political & Comments/replies & -0.257 & 0.000\\
\addlinespace
YouTube football & Number of posts & -0.273 & 0.244\\
YouTube football & Comments/replies & 0.436 & 0.044\\
\addlinespace
YouTube current events & Number of posts & -0.326 & 0.000\\
YouTube current events & Likes/favourites & -0.479 & 0.000\\
YouTube current events & Comments/replies & -0.821 & 0.000\\
\bottomrule
\end{tabular}
\end{table}

\begin{table}[ht]
\caption{\textbf{Mann--Kendall trend statistics for the inverse coefficient of variation.} Most values of Kendall's $\tau$ remain close to $0$, indicating the absence of strong monotonic temporal trends in the concentration levels of the interaction distributions. This suggests that the degree of dispersion and heavy-tailedness remains relatively stable over time across most datasets and interaction types.}
\label{tab:mk_icv}
\centering
\begin{tabular}{llrr}
\toprule
Dataset & Variable & $\tau$ & p-value (sl)\\
\midrule
Gab & Number of posts & -0.213 & 0.065\\
Gab & Comments/replies & -0.153 & 0.187\\
Gab & Retweets/reblogs & -0.204 & 0.077\\
Gab & Likes/favourites & -0.093 & 0.425\\
\addlinespace
X got & Number of posts & -0.181 & 0.224\\
X got & Likes/favourites & -0.147 & 0.315\\
X got & Retweets/reblogs & -0.029 & 0.862\\
X got & Comments/replies & -0.200 & 0.168\\
\addlinespace
X nasa & Number of posts & -0.071 & 0.433\\
X nasa & Likes/favourites & 0.108 & 0.220\\
X nasa & Retweets/reblogs & 0.070 & 0.445\\
X nasa & Comments/replies & 0.099 & 0.264\\
\addlinespace
X cop26 & Number of posts & -0.136 & 0.256\\
X cop26 & Likes/favourites & -0.328 & 0.006\\
X cop26 & Retweets/reblogs & 0.378 & 0.001\\
X cop26 & Comments/replies & -0.180 & 0.132\\
\addlinespace
X ukraine & Number of posts & 0.242 & 0.002\\
X ukraine & Likes/favourites & 0.173 & 0.031\\
X ukraine & Retweets/reblogs & 0.415 & 0.000\\
X ukraine & Comments/replies & -0.002 & 0.981\\
\addlinespace
YouTube political & Number of posts & -0.560 & 0.000\\
YouTube political & Comments/replies & -0.129 & 0.000\\
\addlinespace
YouTube football & Number of posts & 0.385 & 0.077\\
YouTube football & Comments/replies & 0.487 & 0.024\\
\addlinespace
YouTube current events & Number of posts & -0.123 & 0.000\\
YouTube current events & Likes/favourites & -0.274 & 0.000\\
YouTube current events & Comments/replies & -0.120 & 0.000\\
\bottomrule
\end{tabular}
\end{table}

\begin{table}[ht]
\caption{\textbf{Mann--Kendall trend statistics for the log-transformed Gini coefficient.} Most values of Kendall's $\tau$ remain relatively close to $0$, indicating limited monotonic temporal variation in concentration levels. Positive values correspond to increasing concentration over time, while negative values indicate a gradual reduction in inequality. Although some datasets exhibit statistically significant trends, the overall results suggest that high concentration remains a persistent structural characteristic across platforms and interaction types.}
\label{tab:mk_loggini}
\centering
\begin{tabular}{llrr}
\toprule
Dataset & Variable & $\tau$ & p-value (sl)\\
\midrule
Gab & Number of posts & 0.438 & 0.000\\
Gab & Comments/replies & 0.414 & 0.000\\
Gab & Retweets/reblogs & 0.051 & 0.666\\
Gab & Likes/favourites & -0.003 & 0.990\\
\addlinespace
X got & Number of posts & -0.387 & 0.007\\
X got & Likes/favourites & 0.387 & 0.007\\
X got & Retweets/reblogs & 0.080 & 0.602\\
X got & Comments/replies & 0.480 & 0.001\\
\addlinespace
X nasa & Number of posts & -0.223 & 0.011\\
X nasa & Likes/favourites & 0.193 & 0.028\\
X nasa & Retweets/reblogs & 0.256 & 0.005\\
X nasa & Comments/replies & 0.202 & 0.023\\
\addlinespace
X cop26 & Number of posts & 0.627 & 0.000\\
X cop26 & Likes/favourites & 0.254 & 0.033\\
X cop26 & Retweets/reblogs & 0.032 & 0.798\\
X cop26 & Comments/replies & 0.150 & 0.211\\
\addlinespace
X ukraine & Number of posts & -0.269 & 0.001\\
X ukraine & Likes/favourites & 0.141 & 0.079\\
X ukraine & Retweets/reblogs & 0.082 & 0.306\\
X ukraine & Comments/replies & -0.196 & 0.014\\
\addlinespace
YouTube political & Number of posts & 0.693 & 0.000\\
YouTube political & Comments/replies & -0.079 & 0.000\\
\addlinespace
YouTube football & Number of posts & -0.179 & 0.428\\
YouTube football & Comments/replies & -0.333 & 0.127\\
\addlinespace
YouTube current events & Number of posts & 0.644 & 0.000\\
YouTube current events & Likes/favourites & -0.282 & 0.000\\
YouTube current events & Comments/replies & -0.183 & 0.000\\
\bottomrule
\end{tabular}
\end{table}

\end{document}